# Analysis of two Binomial Proportions in Non-inferiority Confirmatory Trials

**Hassan Lakkis[1] and Andrew Lakkis[2,*]**

[1] Biostatistics, Intra-Cellular Therapies , New York, NY, USA
[2] Department of Statistics, Rutgers University, New Brunswick, NJ , USA

**Abstract**
In this paper we propose considering an exact likelihood score (ELS) test for non-inferiority comparison and we derive its test-based confidence interval for the difference between two independent binomial proportions. The p-value for this test is obtained by using exact binomial probabilities with the nuisance parameter being replaced by its restricted maximum likelihood estimate. Calculated type I errors revealed that the proposed ELS method has important advantages for non-inferiority comparisons over popular asymptotic methods for adequately powered confirmatory clinical trials, at 80% or 90% statistical power. For unbalanced sample sizes of the compared treatment groups, type I errors for the asymptotic score method were shown to be higher than the nominal level in a systematic pattern over a range of the true proportions, but the ELS method did not suffer from such a problem. On average the true type I error of the ELS method was closer to the nominal level than all considered methods in the empirical comparisons. Also, in rare cases the type I errors of the ELS test exceeded the nominal level, but only by a small amount. In addition, the p-value and confidence interval using the ELS method can be obtained in less than 30 seconds of computer time for most confirmatory trials. The theoretical arguments and the attractive empirical evidence along with fast computation time should make the ELS method very attractive for consideration in statistical practice.

* Correspondence: Andrew Lakkis, Department of Statistics, Rutgers University, New Brunswick, NJ, USA. Email: andrewlakis56@gmail.com

## 1. INTRODUCTION

It is of interest in confirmatory clinical trials to show whether a new treatment is non-inferior or equivalent to a standard existing treatment. This is very common in therapeutic areas with well-established efficacious treatments. Clinical trials in the anti-infective therapeutic area are frequently designed to show that a new possibly safer or more tolerable drug is non-inferior in treating a specific class of bacterial infections to another active comparator. Common infections that are targeted in these non-inferiority confirmatory clinical trials include, but are not limited to, community acquired pneumonia, complicated skin and skin structure infections, complicated urinary tract infections, and hospital or ventilator acquired pneumonia.

In each of these four diseases, it is not ethical to perform placebo-controlled trials due to the seriousness of the infections and the availability of effective treatments. Non-inferiority designs are also common in vaccine clinical trials for which it is of interest to administer multiple vaccines during a single visit to the clinic. It is of interest to assess whether the conversion in the viral antibody level for each vaccine component in the concomitant administration as non-inferior to that produced from the corresponding vaccine given alone. Travelers to poor areas may be vulnerable to many viral infections and they may need to get their multiple vaccinations concomitantly just before making their trip. Also, pediatric vaccines may need to be administered concomitantly to bring the children up to date in their vaccination schedule.

When designing trials to show that the proportion of responders to the investigational drug is non-inferior to that of the comparator, the clinical trialist must choose a test statistic or a confidence interval method for the difference of proportions that will be used for the non-inferiority test. Several popular methods are available based on asymptotic normality. Researchers must be careful in their consideration of the appropriate method to avoid selecting a method that is excessively conservative or anti-conservative. An anti-conservative test does to not preserve the type I error at the nominal level and the validity of its statistical significance can be questioned.

Common asymptotic methods for confidence interval of difference in two binomial proportions are known to be anti-conservative (liberal) especially when the proportions are higher than 80%. To avoid using anti-conservative confidence intervals, methods based on continuity corrections were proposed by Hauck and Anderson[1] and Newcombe[2]. These methods preserve the type I error below the nominal error rate for most cases, but with frequently reduced power.

Various methods based on exact binomial probabilities were proposed to preserve type I error below the nominal rate. Santner and Snell[3] proposed the unconditional exact method by inverting two one-sided tests based on statistics for the difference between the observed proportions. Chan and Zhang[4] showed that the confidence interval method of Santner and Snell[3] leads to a very conservative confidence interval due to severe discreteness of the distribution of the difference in observed proportions being used, especially for small samples. An unconditional exact non-inferiority test based on the score statistics was proposed by Chan[5] for the p-value, this is known as the exact score (ES) method. Further, Chan and Zhang[4] proposed inverting the two one-sided unconditional ES tests based on the score statistics to obtain the corresponding confidence interval for non-inferiority. The purpose of this exact confidence interval method is that it retains the size of the type I error in each side of the test at below the nominal value regardless of the true value of the nuisance parameter, it is much more efficient than that of Santner and Snell[3] as it is not as conservative, and it is available in SAS 9.3 release[6]. However, in the calculation of the p-value of this ES method the worst-case scenario for the nuisance parameter is assumed, which could be far from the nuisance parameter's likely true value for the study being considered. This could lead to a more conservative test than what would be necessary and a method that utilize the ML value of the true nuisance parameter in calculating the binomial probabilities can lead to a more appropriate test for confirmatory trials. In addition, the ES method require excessive computation time for studies with sample sizes such as those used for Phase 3 confirmatory clinical trials.

Another exact test for non-inferiority comparison was proposed by Wang[7,8]. It was reported by Shan and Wang[9] that a practical issue with this method is when the total sample size for a binomial study includes more than 100 subjects, one would expect an excessive long computation time for obtaining its two-sided confidence interval, even it requires much more computation time to generate the confidence interval than that based on the ES method. For this reason along with that it does not necessarily have higher statistical power than the ES method (based on examples presented by Garner[10]), this exact method will not be included in the remaining part of this article.

Asymptotic methods used for non-inferiority hypothesis tests are usually based on confidence interval for difference of binomial response rates which include: the standard asymptotic normal method known as the Wald's method, the method proposed by Miettinen and Nurminen[11], the uncorrected for continuity method proposed by Newcombe[2], and the method proposed by Agresti and Caffo[12]. The method of Miettinen and Nurminen[11] is in fact the same as the asymptotic score approach as described by Farrington and Manning[13] for non-inferiority hypothesis testing except for bias correction in the variance estimate which has almost no impact on the results when the sample size is large, such as greater than 100 subjects per treatment group which is of interest in this article.

Dann and Koch[14] compared the statistical properties of the 4 methods in the previous paragraph along with the Newcombe[2] with continuity correction method for studies powered at 85%. Based on their results, they recommended the use of each method for various scenarios. Rothmann et al[15], provided an extensive summary of the literature for various methods with applications for non-inferiority comparisons. Kang and Chen[16] proposed an approximate exact score test statistic using the restricted maximum likelihood estimate of the nuisance parameter in the exact binomial probabilities and assessed its properties only up to a total of 100 subjects per treatment group. However, the properties of this test statistic were not assessed for common sample sizes considered in adequately powered clinical trials when the non-inferiority margin is 10%, 15%. In this article, we consider the same ELS test statistic of Kang and Chen[16] for non-inferiority comparisons in difference between investigational and comparator proportions. We also study the properties of this non-inferiority test versus various popular methods for studies with sample sizes selected at statistical powers of 80% or 90% and with non-inferiority margins of 10% or 15% and to a lesser extent with a margin of 5%. In addition, we propose an approach for obtaining a 2 1-sided test-based confidence interval that is consistent with the p-value of this ELS test. The term consistent is utilized to suggest the 1-sided p-value $< \alpha/2$ if and only if the lower side of the 2-sided $100(1-\alpha)$ confidence interval is higher than the non-inferiority margin, $-\delta_0$.

The non-inferiority margins of interest for assessment of statistical properties in this article are 10% and 15%. These are the margins of interest because non-inferiority clinical trials are frequently conducted based on non-inferiority margins ranging from 10% to 15%. Non-inferiority margins are justified by various historical data for effective comparators or agreed on with regulatory agencies prior to study initiation. For instance, each of the US Food and Drug Administration (FDA) guidance documents for the indications of acute bacterial skin and skin structure infections[17], complicated urinary tract infections[18], and complicated intra-abdominal infections[19] reported that a 10% non-inferiority margin is justified by historical evidence for clinical response endpoints. Also, the FDA guidance document for community-acquired bacterial pneumonia[20] reported that a 12.5% non-inferiority margin is justified by historical evidence for a clinical recovery response endpoint. Each of the FDA guidance documents[17,18,19,20] encourages sponsors to discuss the selection of a non-inferiority margin with the FDA in advance of trial initiation, particularly for a proposed margin greater than the ones supported by historical data in the respective guidance documents. Also, assessment of the methods is performed with a non-inferiority margin of 5% when the proportion of response in the comparator is 95% because it is logical to select a smaller margin as the response in the comparator is expected to be closer to 100%.

## 2. BACKGROUND

Assume a binomial trial that compares two groups of sizes $N_T$ and $N_C$. Let X and Y be the independent responses distributed as binomial random variables with parameters $(N_T, P_T)$ and $(N_C, P_C)$, respectively. The test of interest is regarding the difference $(P_T - P_C)$, where this difference is in the range of [-1, 1].

Throughout this article, for simplicity in the hypothesis formulation, we consider that a higher response indicates a better efficacy outcome such as a clinical cure rate in an anti-infective trial. The primary interest is to demonstrate that the new treatment is non-inferior to the standard treatment by a pre-specified margin, such as 10% and 15%. The null hypothesis can be specified as being that the proportion of responders to the new treatment is lower than that of the standard treatment by at least a pre-specified small margin versus the alternative that the proportion of responders to the new treatment is not lower than that of the standard treatment by that margin. Let $P_T$ and $P_C$ denote the true proportions of responders to the test (investigational) and to the control treatments, respectively. Then the test of non-inferiority (one-sided) focuses on the hypothesis:

$$H_0: P_T - P_C \leq -\delta_0 \quad versus \quad H_1: P_T - P_C > -\delta_0 \qquad (1)$$

where $\delta_0$ is the pre-specified non-inferiority margin, a positive quantity. Other articles that discussed the selection of the non-inferiority margin include Temple[21] and the ICH E10 Guidelines[22].

### 2.1. Asymptotic Score Test

To test the non-inferiority hypothesis in (1), the following test statistic can be used:

$$Z(X_T, X_C; \delta_0) = (\hat{P}_T - \hat{P}_C + \delta_0)/\sqrt{\tilde{v}_0} \qquad (2)$$

where $\hat{P}_T$ and $\hat{P}_C$ are the observed proportions of responders in the two groups, respectively, and $v_0 = P_T(1 - P_T)/N_T + P_C(1 - P_C)/N_C$ is the variance of the numerator and $\tilde{v}_0$ is its maximum likelihood (ML) estimate obtained by replacing $P_T$ and $P_C$ by their ML estimates under the constrained null hypothesis of $P_T - P_C = -\delta_0$. Large values of the test statistic favor the alternative hypothesis. The test statistic provided in (2) will be referred to as the asymptotic score test statistic when $Z(X_T, X_C; \delta_0)$ is considered to follow the asymptotic standard normal distribution. Its p-value can be obtained by using

$$p_{\delta_0}(x_T, x_C) = Pr(Z(X_T, X_C; \delta_0) \geq Z(x_T, x_C; \delta_0)|\delta_0) \qquad (3).$$

The test-based 2-sided confidence interval that corresponds to (2) can be obtained by inverting the following equation for $\delta$

$$Z_{\alpha/2}^2 = (\hat{P}_T - \hat{P}_C + \delta)^2 / (\tilde{P}_T(1 - \tilde{P}_T)/N_T + \tilde{P}_C(1 - \tilde{P}_C)/N_C) \qquad (4)$$

where $\tilde{P}_T$ and $\tilde{P}_C$ are the ML estimates under the constraint $P_T - P_C = -\delta$ and $Z_{\alpha/2}$ is the upper $\alpha/2$ percentile of the standard normal distribution. This is in fact an asymptotic likelihood score method and will be referred as the ALS method. Miettinen and Nurminen (MN)[11] proposed this approach to produce the confidence interval given in (4) and provided expressions for the ML estimates $\tilde{P}_T$ and $\tilde{P}_C$, but multiplied the denominator by $N/(N-1)$ with $N = N_T + N_C$ to correct for bias in the variance estimate. Since the focus of this article is on large sample sizes, correcting the denominator of (4) by $N/(N-1)$ would not have meaningful impact and therefore we drop this factor for the remainder of this article. Farrington and Manning[13] proposed the score test statistic given in (2) and provided an estimate of the sample size for a given power.

### 2.2. Exact Score Test

Chan[5] proposed using the exact unconditional binomial distribution of the test statistics $Z(X_T, X_C; \delta_0)$ for all possible outcomes of the two binomial responses, where each outcome $(X_T, X_C)$ corresponds to a 2×2 table. Following his approach, the probability of observing a particular outcome $(x_T, x_C)$ is a product of two binomial probabilities. Let $P = P_T$, the likelihood for the test of non-inferiority given in (1) can be written as

$$Pr(X_T = x_T, X_C = x_C | H_0) = \binom{N_T}{x_T}\binom{N_C}{x_C} P^{x_T}(1-P)^{N_T-x_T}(P+\delta_0)^{x_C}(1-P-\delta_0)^{N_C-x_C} \quad (5)$$

Let $Z(x_T, x_C; \delta_0)$ be the value of $Z(X_T, X_C; \delta_0)$ at the observed $(x_T, x_C)$ table using equation (2). As described by Chan[5], the exact significance level is the sum of the probabilities of tables that are at least as extreme as the observed table. As can be seen, the likelihood in (5) depends on an unknown nuisance parameter $P$. The exact significance level for the test of the hypothesis in (1) is calculated by maximizing the null likelihood over the domain of the nuisance parameter $P$. The domain of $P$ is $D = [0, 1-\delta_0]$. The exact significance level or p-value for the score test is given by

$$p_{\delta_0}(x_T, x_C) = \max_{P \in D} \Pr(Z(X_T, X_C; \delta_0) \geq Z(x_T, x_C; \delta_0) | \delta_0, P) \quad (6)$$

This is the ES test, which was proposed by Chan[5] with p-value as given in (6), is an extension of the work of Suissa and Shuster[19] to include null hypotheses of a pre-specified difference that is different from 0.

Chan[5] provided detailed steps describing how to calculate the p-value given in (6). To obtain adequate accuracy of the p-value, Chan[5] suggested dividing the domain of $P$ into 1,000 equally spaced subintervals and at each value of $P$ calculate the probability on the right side of equation (6). The p-value is the maximum of these 1,000 probabilities. Chan and Zhang[4] derived the exact test-based confidence interval by inverting the exact score test statistic.

Due to the numerical search over the domain of the nuisance parameter to obtain the p-value given in (6), this computation process requires intensive computer work as the sample size increases. Chan[5] indicated that calculating the corresponding test-based confidence interval is much more complicated than performing the hypothesis test because it requires numerical search in the parameter space for the nuisance parameter as well as for the parameter space for each bound of the confidence interval.

### 2.3. Exact Likelihood Score Test

This ELS test described in this paragraph is the one of primary interest in this article. Based on the same approach of the ES test we will follow the approach of replacing the nuisance parameter $P$ in (6) by its ML estimate, $\tilde{P}$, under the constraint $P_T - P_C = -\delta_0$. Then, the p-value for this ELS test become much simpler to obtain and can be written as

$$p_{\delta_0}(x_T, x_C) = \Pr(Z(X_T, X_C; \delta_0) \geq Z(x_T, x_C; \delta_0) | \delta_0, \tilde{P}) \quad (7)$$

This p-value given in (7) can be obtained by following the 2 steps:

1. Calculate the $Z(X_T, X_C; \delta_0)$ statistic for all possible binomial outcomes and order them.

2.  The p-value in (7) can be obtained by summing up all of the probabilities using the likelihood function in (5), with $P$ being replaced by $\tilde{P}$, for the outcomes whose $Z(X_T, X_C; \delta_0)$ statistic is greater than or equal to the observed $Z(x_T, x_C; \delta_0)$ statistic at the $(x_T, x_C)$ table.

Storer and Kim[24] proposed the ELS test in (7) by obtaining the p-value using the maximum likelihood estimate of the nuisance parameter for testing the case of zero difference under the null hypothesis between the binomial proportions. Also, they discussed the attractive properties of the ELS test relative to the exact method of Suissa and Shuster[23] by showing that the type I error of their method is closer to the nominal rate and required much less computation time. Kang and Chen[16] extended the ELS test to the null hypothesis of non-zero difference, based on the same p-value given in (7), for non-inferiority testing. However as stated earlier, they limited their assessment of the statistical properties to a maximum of 100 subjects per treatment group. In this article, we assess the statistical properties of this method versus other commonly used methods for sample sizes that are required for adequately powered studies for non-inferiority confirmatory trials with non-inferiority margins 15%, 10%, and to a less extent 5%. Most confirmatory trials have treatment allocations ratio of $N_T: N_C$ =1:1, 2:1, 1:2 leading to sample sizes frequently being in the range from 100 to over 600 subjects per treatment group.

In addition to the p-value given in (7) we discuss obtaining a consistent confidence interval which is necessary for reporting non-inferiority hypothesis test results for confirmatory trials.

### 3. PROPOSED CONFIDENCE INTERVAL FOR THE NON-INFERIORITY TEST

For presentation of results for non-inferiority clinical trials, clinical reports customarily include confidence intervals along with the p-values as the basis for clinical interpretation and for non-inferiority statistical inference. However, the FDA has been reporting the confidence intervals as the basis for non-inferiority confirmatory results in Package Inserts for approved products without inclusion of the p-values. Accordingly, to make the ELS method more appealing for consideration in clinical trial practice, we propose deriving a test-based confidence interval that is consistent with the p-value given in (7).

When selecting the bounds of the confidence interval, we would like to preserve the decision rule that is consistent with the p-value for the hypothesis test. Let $\delta_L$ and $\delta_U$ represent the lower and upper bounds of the confidence interval for $\delta$. The lower bound is obtained by inverting the one-sided, $\alpha/2$, test of hypothesis for $H_0$: $P_T - P_C = \delta_0$ versus $H_1$: $P_T - P_C > \delta_o$ and the upper bound is obtained by inverting the one-sided, $\alpha/2$, test of hypothesis for $H_0$: $P_T - P_C = \delta_0$ versus $H_1$: $P_T - P_C < \delta_o$. Therefore, the $100(1-\alpha)\%$ confidence interval $(\delta_L, \delta_U)$ for the difference $\delta = P_T - P_C$ can be obtained using:

$$\delta_L = \inf_{\delta}\{\delta: \Pr[Z(X_T, X_C; \delta) \geq Z(x_T, x_C; \delta)|\delta, \tilde{P}] > \alpha/2\} \quad (8)$$

$$\delta_U = \sup_{\delta}\{\delta: \Pr[Z(X_T, X_C; \delta) \leq Z(x_T, x_C; \delta)|\delta, \tilde{P}] > \alpha/2\} \quad (9)$$

where the probabilities in (8) and (9) are evaluated using the exact binomial distributions as presented in (5) with $P$ being replaced by $\tilde{P}$. Chen[21] proposed deriving a confidence interval using the maximum likelihood estimate of $P$ in the binomial probabilities by inverting the two sided hypothesis $H_0$: $P_T - P_C = \delta_0$ versus $H_1$: $P_T - P_C \neq \delta_o$ based on the score statistic given in (2). Chen's method was proposed with focus on minimizing the width of the confidence interval and its lower bound cannot be used for non-inferiority hypothesis testing since the size of the lower bound one-sided test is not fixed at $\alpha/2$.

It should be noted that each of the regions

$$I(\delta) = \{\delta: \Pr[Z(X_T, X_C; \delta) \geq Z(x_T, x_C; \delta) | \delta, \tilde{P}] > \alpha/2\} \quad (10)$$
$$J(\delta) = \{\delta: \Pr[Z(X_T, X_C; \delta) \leq Z(x_T, x_C; \delta) | \delta, \tilde{P}] > \alpha/2\} \quad (11)$$

could be a union of disjointed intervals. Chen[25] as well as Chan and Zhang[4] also encountered the same problem of disjoint intervals for similarly defined regions of (10) and (11) when proposing the confidence limits for the difference of binomial proportions using the score statistics. This problem presents a significant computation challenge to estimate $(\delta_L, \delta_U)$ correctly based on (8) and (9) by requiring very extensive computation, leading to significant delay in computer time to generate the estimates of the confidence limits.

To simplify computations of the confidence limits based on the score statistics, Chen[25] suggested using the bisection method and pointed out that in practice when expressions in his article, similar to (8) or (9), provide multiple solutions they tend to be very close to each other and their difference has negligible impact on the coverage probability of the true difference in proportions.

Instead of using the formula in (8) and (9), we propose building the confidence interval based on the following two probabilities denoted by

$$g_L(\delta) = \Pr[Z(X_T, X_C; \delta_{LSC}) \geq Z(x_T, x_C; \delta_{LSC}) | \delta, \tilde{P}] \quad (12)$$
$$g_U(\delta) = \Pr[Z(X_T, X_C; \delta_{USC}) \leq Z(x_T, x_C; \delta_{USC}) | \delta, \tilde{P}] \quad (13)$$

where $(\delta_{LSC}, \delta_{USC})$ are the lower and upper bounds of the confidence interval based on the asymptotic score method given in (4). Note that defining the rejection regions as $[Z(X_T, X_C; \delta_{LSC}) \geq Z(x_T, x_C; \delta_{LSC}) | \tilde{P}]$ and $[Z(X_T, X_C; \delta_{LSC}) \leq Z(x_T, x_C; \delta_{LSC}) | \tilde{P}]$ for (12) and (13), respectively, to be independent of $\delta$ causes the probabilities $g_L(\delta)$ and $g_U(\delta)$ to be polynomials in $\delta$ and therefore both are continuous functions in $\delta$. This same approach of modifying the critical region to be independent of $\delta$ was considered by Chan and Zhang[4] for obtaining the exact confidence interval based on the unconditional exact score method. The lower and upper bounds of the confidence interval $(\delta_L, \delta_U)$ for the ELS method based on (12) and (13) can be found by solving $g_L(\delta) = \alpha/2$ and $g_U(\delta) = \alpha/2$ for $\delta$, respectively, where the range of $\delta$ is $(-1, 1)$. The lower bound $\delta_L$ can be obtained by solving $g_L(\delta) = \alpha/2$ for $\delta$ using the bisection method with the lower bound of the asymptotic score method as one of the initial values. Similarly, the upper bound $\delta_U$ can be obtained by solving $g_U(\delta) = \alpha/2$ for $\delta$ using the bisection method with the upper bound of the asymptotic score method as one of the initial values. The resulting $(\delta_L, \delta_U)$ confidence interval is consistent with the hypothesis test p-value given in (7).

With the improvement of the speed of PC computers up to present time, the suggested method for obtaining the confidence interval and the p-value as described above can be calculated in less than 30 seconds of real computer time even for studies with total sample size in the two compared groups of 1000 subjects. As examples, we provide the real computer time for obtaining the proposed confidence interval and the p-value in Section 7.

To assess usefulness of the ELS method for confirmatory trials, we consider comparing its statistical properties to various common methods in the literature based on studies with adequate sample sizes for hypotheses testing with statistical powers of 80% and 90%. The next two sections describe the methods that will be included in the assessment along with the statistical properties being used.

## 4. TESTS COMPARED

Based on statistical properties we compare the ELS method to six common asymptotic methods for testing non-inferiority hypotheses based on two independent proportions.
The methods considered in the comparison of the statistical properties include the followings:

1) The ELS method with its p-value as presented earlier in (7) and its confidence interval as describe in Section 3 in this article.

2) The commonly known Wald method whose confidence interval is given by:

$$(\hat{P}_T - \hat{P}_C) \pm Z_{\alpha/2}\sqrt{\hat{P}_T(1-\hat{P}_T)/N_T + \hat{P}_C(1-\hat{P}_C)/N_C} \qquad (14)$$

Where $Z_{\alpha/2}$ is the cut-off from the standard normal distribution.

3) The Agresti-Caffo (AC)[12] method whose confidence interval is given by

$$(\bar{P}_T - \bar{P}_C) \pm Z_{\alpha/2}\sqrt{\bar{P}_T(1-\bar{P}_T)/(N_T+2) + \bar{P}_C(1-\bar{P}_C)/(N_C+2)} \qquad (15)$$

where $\bar{P}_T = (x_T + 1)/(N_T + 2)$ and $\bar{P}_C = (x_C + 1)/(N_C + 2)$. In essence, the AC method simply adds one success and one failure to each treatment group and then uses the same Wald's formula for the confidence interval. This simple adjustment gives this method significant advantages over the Wald's method, as was shown by AC[12] and as will be shown later in this article.

4) The Newcombe Hybrid Score interval (NC)[2] whose bounds are given by

$$\text{Lower Bound:} \quad (\hat{P}_T - \hat{P}_C) - \sqrt{(\hat{P}_T - l_T)^2 + (u_C - \hat{P}_C)^2} \qquad (16)$$
$$\text{Upper Bound:} \quad (\hat{P}_T - \hat{P}_C) + \sqrt{(u_T - \hat{P}_T)^2 + (\hat{P}_C - l_C)^2} \qquad (17)$$

where $l_T$ and $u_T$ are the lower and upper limits of the Wilson Score interval for $P_T$ and are obtained by solving the following equation for $P_T$

$$|P_T - \hat{P}_T| = Z_{\alpha/2}\sqrt{P_T(1-P_T)/N_T}. \qquad (18)$$

Similarly, $l_C$ and $u_C$ are the lower and upper limits of the Wilson Score interval for $P_C$ and can be obtained by solving a similar score equation to (18) for $P_C$.

5) The asymptotic likelihood score confidence interval method (ALS) as given in (4) and its test statistic as given in (2). This is the same as the Miettinen and Nurminen[11] method without the bias correction in the denominator. Also, a Farrington-Manning[13] interval which can lead to the same statistical significance as it is derived from the same test statistic and can be obtained by

$$(\hat{P}_T - \hat{P}_C) \pm Z_{\alpha/2}\sqrt{\tilde{P}_T(1-\tilde{P}_T)/N_T + \tilde{P}_C(1-\tilde{P}_C)/N_C} \qquad (19)$$

where $\tilde{P}_T$ and $\tilde{P}_C$ are the maximum likelihood estimates of $P_T$ and $P_C$, respectively, calculated under the non-inferiority null hypothesis constraint of $P_T - P_C = -\delta_0$.

All three confidence interval methods are test-based and derived from the same asymptotic score test statistic as given in (2). Therefore, all three methods will have the same statistical significance and shall be referred to as the asymptotic score method in the comparisons.

6) The Hauck-Anderson continuity (HA)[1] corrected interval. Its bounds are given by:

$$(\hat{P}_T - \hat{P}_C) \pm \left\{ Z_{\alpha/2} \sqrt{\hat{P}_T(1 - \hat{P}_T)/N_T + \hat{P}_C(1 - \hat{P}_C)/N_C} + 1/[2 \min(N_T, N_C)] \right\}. \quad (20)$$

7) The Newcombe Hybrid Score continuity corrected interval (NCC). Newcombe[2] provided this continuity corrected method by solving $|P_T - \hat{P}_T| - 1/2N_T = Z_{\alpha/2}\sqrt{P_T(1 - P_T)/N_T}$ and $|P_C - \hat{P}_C| - 1/2N_C = Z_{\alpha/2}\sqrt{P_C(1 - P_C)/N_C}$ for the confidence limits of $P_T$ and $P_C$. Newcombe then substituted the appropriate confidence limits of $P_T$ and $P_C$ in the interval given in (16) and (17).

## 5. EXACT TYPE I ERRORS

The statistical methods presented in Section 4 for testing non-inferiority hypotheses will be compared based on the statistical criteria of the true type I error.

For the test-based confidence interval methods (ELS and ALS methods), we define the rejection region of an arbitrary level $\alpha/2$ one-sided test with non-inferiority margin $\delta_0$ as follows:

$$R_0 = \{(i,j): (i,j) \in \Omega \text{ and } p_{\delta_0}(i,j) \leq \alpha/2\} \quad (21)$$

where $\Omega = \{0,1,\ldots,N_T\} \times \{0,1,\ldots,N_C\}$ represents the entire sample space for the multivariate random variable $(X_T, X_C)$, and $p_{\delta_0}(i,j)$ is the one-sided p-value for the non-inferiority test, calculated as described in (3) for ALS method and in (7) for the ELS method. This p-value is calculated for each possible observation $(i,j)$ in the entire sample space $\Omega$ to form the critical region.

The other tests in the comparisons included the Wald, AC, NC, NCC, and HA were based on comparing the lower bounds of their confidence intervals versus $-\delta_0$ and these bounds were not inverted from test statistics. Based on each of these methods a lower bound of the confidence interval greater than $-\delta_0$ implies rejection of the null hypothesis and non-inferiority is concluded. For these methods, the rejection region of a level $\alpha/2$ one-sided test can be defined as follows:

$$R_0 = \{(i,j): (i,j) \in \Omega \text{ and } \delta_l(i,j) > -\delta_0\} \quad (22)$$

where $\delta_l(i,j)$ is the lower bound of the 2-sided $100(1 - \alpha)\%$ confidence interval for a given method at the $(i,j)$ realization of $(X_T, X_C)$ in the overall domain $\Omega$. Each side outside the 2-sided confidence interval has $\alpha/2$ probability.

For given sample sizes $(N_T, N_C)$ and $P_C$, the exact type I error for each of the methods for testing $H_0: P_T - P_C = -\delta_0$ versus $H_1: P_T - P_C > -\delta_0$ is given by $\alpha_{\delta_0}(P) = P_{H_0}\{(X_T, X_C) \in R_0|P\}$ and can be obtained using the following equation:

$$\alpha_{\delta_0}(P) = \sum_{(i,j) \in R_0} \binom{N_T}{i}\binom{N_C}{j} P^i (1-P)^{N_T-i}(P+\delta_0)^j(1-P-\delta_0)^{N_C-j} \quad (23)$$

where $P = P_T = P_C - \delta_0$.

## 6. RESULTS

Summary comparisons of the exact type I error rates calculated based on the formula given in (23) for the methods summarized in Section 4 for the test of non-inferiority versus the one-sided 0.025 nominal value level are presented in Tables 1 and 2 to assess the performance of the ELS method versus the other asymptotic methods. The sample sizes used in Tables 1 and 2 were generated using SAS program based on 80% and 90% statistical power, respectively, using the Farrington and Manning[13] method to show non-inferiority based on various scenarios of the parameters (non-inferiority margin, sample size ratio of test:control, and proportion of response in the control group) for the hypothesis test. The true type I error rates were also computed using SAS IML codes for the following combinations: 1) a non-inferiority margin of 10%, a randomization ratio of Test:Control of 1:2, 1:1, and 2:1 and a proportion of response in the control group of 25%, 40%, 60%, 75%, and 90%, and 95%; 2) for the case of the non-inferiority margin of 15%, the same combinations as for non-inferiority margin of 10% are used without the case of proportion of response in the control group of 95%. For the case of the non-inferiority margin of 5%, the true type I error rate is computed in Tables 1 and 2 for the proportion of response in the control group of 95% for the three subject allocation ratios to Test:Control (1:2, 1:1, and 2:1).

### 6.1. Comparison of Type I error

From both Tables 1 and 2, each method's type I error was evaluated based on 72 (36 per table) different combinations of statistical power, randomization ratio, non-inferiority margin, and control proportion. For easy identification, type I errors in Tables 1 and 2 higher than the 2.5% nominal value are in bold fonts and those below 2.0% are also in bold fonts. Table 3 summarized the type I errors across the various scenarios presented in Tables 1 and 2.

Comparisons of the asymptotic methods: the empirical evidence based on the type I errors for the 72 scenarios summarized in Table 3 and presented in Tables 1 and 2 confirmed that the ALS method has the best properties in term of having its type I error closest to the 2.5% nominal level on average versus all other asymptotic methods. In comparison to the ALS method: 1) the NC and AC methods had type I errors that were very frequently higher than the 2.5% nominal level in 69% and 74%, respectively, of the 72 cases; 2) the Wald's method had wide range of the type I errors (1.98% to 4.03%) and 49% of them were higher than the 2.5% nominal level versus a narrower range (2.05% to 2.84%) of those of the ALS method and with 42% of them being higher than the 2.5% nominal level; 3) the methods with continuity correction of HA and NCC were frequently very conservative with many of their type I errors being much lower than the 2.5% nominal level, making them frequently having less statistical power than the ALS method.

Comparisons of the ALS and the ELS methods: the empirical comparison based on the summary of the type I errors displayed in Table 3 supports that the ELS method outperforms the ALS method based on all displayed measures. Specifically, the type I errors were higher than the 2.5% nominal level in 8% for the ELS method versus 42% for the ALS method, their average distance from the 2.5% nominal level was 0.046 for the ELS method versus 0.120 for the ALS method, they ranged from 2.31% to 2.58% (range=0.27) for the ELS method versus from 2.05% to 2.84% (range= 0.79) for the ALS method, the average of those ≤2.5% was 2.45% for the ELS method versus 2.38% for the ALS method, and the average of those >2.5% was 2.53% for the ELS method versus 2.62% for the ALS method.

In addition, when the scenario for randomization ratio is 1:2 in test:control groups, the type I errors for the ALS method decrease from values systematically higher than 2.5% nominal level to values lower than the 2.5% level (Tables 2 and 3). Moreover, when the scenario for randomization ratio is 2:1 the type I errors increase from values lower than 2.5% to values systematically higher than 2.5% as the proportion of the control group increases. To illustrate this problem we will focus on the empirical data from Table 2 for trials with sample sizes selected at 80% statistical power, for the cases of 2:1 randomization ratio, non-inferiority margin 10%, proportions of the control group: 0.25, 0.40, 0.60, 0.75, 0.90, 0.95: the type I errors for the ALS method (systemically increased beyond 2.5% level) were: 2.39%, 2.47%, 2.49%, **2.59%, 2.69%, 2.82%,** respectively, versus the type I errors for the ELS method: 2.48%, 2.50%, 2.48%, 2.48%, 2.48%, 2.38%, respectively.

For the scenarios of 1:1 randomization ratio, the ALS method controlled the type I error at the 2.5% nominal level in 75% of the cases (18 out of 24 cases) and there was no pattern of systematic lack of control of type I error with various ranges of the proportions in the control group. Based on the scenarios with randomization ratio 1:1, the ELS method controlled the type I errors in 92% of the cases at the 2.5% nominal level, in 22 out of the 24 cases.

## 7. EXAMPLES

Three examples of non-inferiority analysis are presented in Table 4. The methods used for the analysis include the ELS and the six asymptotic methods. SAS IML programs were prepared and used on a MacBook Pro computer to generate the confidence intervals for these methods and the p-values for the ELS and ALS methods. The three examples are not for actual clinical trials data, but the ranges of the sample sizes, the responses, and the non-inferiority margins are similar to those that can appear in anti-infective Phase 3 clinical trials. The real computer time is also reported for each example for obtaining the confidence interval and the p-value for the ALS method. The reported computer time was based on a MacBook Pro computer which had Dual-Core Core i5 processor, 2.3 GHz speed, and 8 GB memory.

*Example 1* included 328 and 317 subjects in the test and the control groups, respectively. The non-inferiority margin was selected at 10% and the observed responses were assumed as 264 and 268 in the test and the control groups, respectively. The sample size of this example was chosen so that the one-sided test has approximately 90% power to rule out the inferiority of the test group to the control group by 10% with α= 2.5%, assuming the response in the control group is 82%. As shown in Table 4, the three methods with bolded confidence intervals imply statistical significance. However, these methods had expected type I errors that is likely to exceed the nominal 2.5% based on true proportions being close to the observed proportions as was shown in Tables 2 and 3. The two methods with continuity corrections are not significant, which suggests that they had lower power than the other methods. Based on the ELS method, the lower bound of the confidence interval is -9.94% which is higher than -10% and that implies statistical significance in rejecting the inferiority margin. Also, this significance based on the confidence interval is consistent with the p-value of 0.0238 which is less than 0.025. The ALS test is significant and almost identical to the proposed ELS method in this example. It should be noted that it took only 12 seconds of real computer time to generate the p-value and the confidence interval based on the ELS method for this example.

*Example 2* included 326 and 108 subjects in the test and the control groups, respectively. This example illustrates results of a study with randomization ratio of approximately 3:1 in the test and control groups, respectively. The non-inferiority margin was assumed at 10% and the observed responses were assumed as 285 and 99 in the test and the control groups, respectively. The sample size of this example was chosen so that the one-sided test has approximately 90% power to rule out inferiority of the test group to the control group by 10% with $\alpha= 2.5\%$, assuming the response in the control group is 90%. As shown in Table 4, the two methods with bolded confidence interval imply statistical significance, but these methods likely have true type I errors based on true proportions being close to the observed ones that exceed the nominal $\alpha= 2.5\%$ (type I errors were generated for various scenarios of the case of 3:1 randomization ratio but not included in this article). All other methods are not significant since the lower bounds of their confidence intervals are below -10%. However, the ELS method is closest to statistical significance, (to -10%), based on the lower bound of the confidence interval. Result based on the p-value for the ELS method did not indicate significance at the 0.025 level and therefore was consistent with the conclusion based on the corresponding 95% confidence interval. It should be noted that for this example it took only 5 seconds of real computer time to generate the p-value and the confidence interval for the ELS method.

*Example 3* included 435 and 441 subjects in the test and the control groups, respectively. This example represents a study with randomization ratio of approximately 1:1 in test and control groups, respectively. The non-inferiority margin was assumed at 5% and the observed responses were assumed as 411 and 426 in the test and the control groups, respectively. The sample size of this example was chosen so that the one-sided tests have approximately 90% power to rule out inferiority of the test group to the control group by 5% with $\alpha= 2.5\%$, assuming the response in the control group is 95%. As shown in Table 4, the two methods with bolded confidence interval led to statistical significance, but their type I errors likely to be higher than the nominal of 2.5% when the true proportions are near the observed ones as shown in Table II. However, the only other methods that imply statistical significance are the HA and the ELS method based on the lower bounds of their confidence intervals. Also, the p-value for the ELS is 0.0246 which is less than 0.025 leading to consistent conclusion to that based on the confidence interval. For this example, it took only 23 seconds of real computer time to generate the p-value and the confidence interval based on the ELS method.

## 8. DISCUSSION

Nurminen and Newcombe[26] in an article published in 2009 concluded that the MN likelihood score interval estimates are highly recommended on account of both theoretical arguments and empirical evidence for use in epidemiology and medical research practice. The MN[11] likelihood score interval method is the same as the ALS method in this article except the denominator of the MN likelihood method included a bias correction factor for the variance estimate, $N/(N-1)$ with $N = N_T + N_C$, which makes no practical difference in confirmatory studies with sample sizes frequently above 150 subjects per treatment group. In fact, our empirical type I errors as presented for the various scenarios in Tables 2 and 3 and summarized in Table 1 confirm their conclusion in terms of the advantage of the ALS method over the other 5 asymptotic methods. The ALS method has average type I errors much closer to the 2.5% nominal level than the two methods with continuity corrections as their type I errors are frequently much lower than the nominal 2.5% level. Also, the ALS method is preferred over the Wald, AC, and the NC methods by having smaller percentage of type I errors being >2.5% nominal level and for those cases of the ALS method with type I errors >2.5% nominal level on average its type I errors stay closer to the nominal 2.5% level.

On a theoretical basis, both the ELS and ALS methods depend on the constrained ML estimates of the proportions of the compared groups in estimating the variance of the binomial probabilities. The only important difference between the two methods is that the ELS method utilizes the exact binomial probabilities with the nuisance parameter being replaced by its constrained ML estimates of the binomial proportions to obtain the p-values and the confidence intervals while ALS method approximates the exact binomial by the asymptotic normal distribution to facilitate its computations without the need to deal with the nuisance parameter. Because the ELS method does not have such asymptotic approximation of the binomial distribution it is expected to have better statistical properties than the ALS method and that was confirmed empirically by the type I errors summarized and displayed in Tables 1, 2, and 3. Based on the results presented in Tables 2 and 3 for trials with sample sizes commonly considered in non-inferiority confirmatory studies, the type I errors for the ELS method were frequently (92%) below or equal to the nominal value of 2.5% than those of the ALS method (58%), had much closer distance to the 2.5% nominal level than those of the ALS method, had narrower range than those of the ALS method, and the cases with type I errors >2.5% nominal level the type I errors for the ELS method stayed closer to the nominal value than those of the ALS method.

The ALS method had a systematic problem of not controlling the type I errors at the 2.5% nominal level with specific scenarios based on the empirical type I errors displayed in Tables 2 and 3. Specifically, the ALS method had type I errors that are systematically higher than the 2.5% nominal level when the number of subjects in the test:control groups had a ratio of 2:1 and the proportion of response in the control group is 60% or higher and this problem increased as the proportion of the control group increased to 95%. In addition, this problem of systematically higher type I errors than the nominal level also appeared for the ALS method for the case of randomization ratio of 1:2 combined with low proportions (25% and 40%) in the control group. Based on other empirical data not presented in this article, we observed that this is a general problem with the ALS method as systematically not controlling the type I errors at the nominal level when the randomization ratio is imbalanced between the compared treatment groups combined with specific range of the proportion of response in the control group. Unlike the ALS method, the type I errors for the ELS method frequently controls the type I errors at the nominal level regardless of randomization ratios and/or of range of the proportions of responses of the treatment groups.

In practice of analysis of confirmatory clinical trials, so far as significance testing is concerned, non-inferiority effect will confidently be considered significant (rejecting inferiority and concluding non-inferiority) only if it remains significant based on sensitivity analysis that corrects for inflated type I error over the nominal level of the primary comparison. As shown earlier, this can affect the ALS score method when the p-value is significant and with value being close to the nominal value, but it is less of a concern for the ELS method.

Unlike other exact methods, the ELS method was not developed to control the type I error at the nominal level at the full domain of the nuisance parameter and therefore it is not as conservative while it preserves the type I error frequently at the nominal level (as shown in 92% of the considered cases in Tables 2 and 3). Availability of the confidence interval for analyzing non-inferiority clinical trial data is very important as it is the primary statistical method for analyzing and reporting non-inferiority tests in clinical reports and product label.

In term of actual computation time to generate the confidence interval and the p-value for the ELS method, it can readily be generated for the difference between two proportions in less than 30 seconds for most studies with 80%-90% statistical power. We also ran the SAS IML program to generate the confidence interval and p-value based examples of larger trials to assess real computer time based on studies with total sample sizes in the range of 2,000-2,100 subjects divided between the two compared groups with randomization ratios of 2:1, 1:1, and 1:2. The real computer time for these examples ranged from 1 minute to 2 minutes suggesting the ELS method is very feasible for consideration in term of computation time for trials with sizes in excess of 2,000 subjects.

We recommend considering the ELS method over the ALS method for the analysis of confirmatory clinical trials and various similar medical research given its theoretical advantage and empirical evidence of being the method with better statistical properties. In addition, given the superior performance of the ELS method we encourage statistical software developers to make this method available in their future updates.

The SAS IML codes used to generate the p-value and the confidence interval for the ELS method and all other methods reported in the examples in this article will be made available upon request from by the corresponding author.

## ACKNOWLEDGMENT

The authors would like to thank Haiyuan Zhu for his helpful comments.

Analysis of two Binomial Proportions in Non-inferiority Trials

Table 1: True Type 1 Errors for Various Tests for Studies with 80% Power at Nominal $\alpha=2.5\%$

| $\delta_0$ | $N_T:N_C$ | $P_C$ | $N_T$ | Wald | AC | HA | NCC | NC | ALS | ELS |
|---|---|---|---|---|---|---|---|---|---|---|
| 0.10 | 1:2 | 0.25 | 207 | 2.21 | 2.45 | **1.83** | 2.31 | **2.77** | **2.65** | 2.49 |
| | | 0.40 | 275 | 2.34 | 2.50 | 2.06 | 2.30 | **2.62** | **2.56** | 2.48 |
| | | 0.60 | 285 | 2.46 | 2.47 | 2.19 | 2.18 | 2.46 | 2.44 | 2.44 |
| | | 0.75 | 233 | **2.70** | **2.60** | 2.34 | 2.13 | 2.48 | 2.44 | 2.47 |
| | | 0.90 | 132 | **3.38** | **2.80** | **2.60** | 1.72 | 2.33 | 2.29 | **2.51** |
| | | 0.95 | 90 | **3.94** | **3.08** | **3.05** | 1.41 | 2.05 | 2.05 | 2.39 |
| | 1:1 | 0.25 | 295 | **2.57** | **2.57** | 2.25 | 2.16 | **2.57** | 2.50 | 2.50 |
| | | 0.40 | 374 | 2.46 | **2.54** | 2.27 | 2.28 | **2.57** | **2.51** | 2.46 |
| | | 0.60 | 374 | 2.37 | 2.43 | 2.36 | 2.36 | 2.50 | 2.40 | 2.38 |
| | | 0.75 | 295 | 2.50 | **2.54** | 2.24 | 2.24 | **2.58** | 2.50 | 2.47 |
| | | 0.90 | 154 | **2.78** | **2.76** | 2.19 | 1.97 | **2.55** | 2.36 | 2.47 |
| | | 0.95 | 99 | **3.27** | **2.91** | 2.47 | 1.67 | 2.42 | 2.32 | 2.32 |
| | 2:1 | 0.25 | 466 | **2.88** | **2.67** | 2.48 | 2.02 | 2.42 | 2.39 | 2.48 |
| | | 0.40 | 570 | **2.62** | **2.57** | 2.31 | 2.20 | **2.52** | 2.47 | 2.50 |
| | | 0.60 | 550 | 2.47 | 2.48 | 2.20 | 2.23 | **2.54** | 2.49 | 2.48 |
| | | 0.75 | 414 | 2.29 | 2.49 | **1.94** | 2.27 | **2.65** | **2.59** | 2.48 |
| | | 0.90 | 194 | 2.13 | **2.52** | **1.52** | 2.16 | **2.97** | **2.69** | 2.48 |
| | | 0.95 | 116 | 2.00 | **2.67** | **1.37** | 1.97 | **2.88** | **2.82** | 2.38 |
| 0.15 | 1:2 | 0.25 | 90 | 2.14 | **2.56** | **1.49** | 2.25 | **2.94** | **2.57** | 2.46 |
| | | 0.40 | 120 | 2.22 | 2.48 | **1.81** | 2.26 | **2.76** | **2.63** | 2.48 |
| | | 0.60 | 127 | **2.61** | **2.61** | 2.12 | 2.20 | **2.61** | **2.61** | 2.47 |
| | | 0.75 | 106 | **2.76** | **2.67** | 2.22 | 2.03 | **2.53** | 2.46 | 2.50 |
| | | 0.90 | 65 | **3.55** | **2.87** | **2.56** | 1.65 | 2.22 | 2.19 | 2.51 |
| | 1:1 | 0.25 | 131 | **2.75** | **2.77** | 2.14 | **1.96** | 2.56 | 2.37 | 2.50 |
| | | 0.40 | 165 | 2.47 | **2.63** | 2.11 | 2.16 | **2.66** | **2.58** | **2.52** |
| | | 0.60 | 165 | 2.38 | **2.76** | 2.13 | 2.13 | **2.76** | **2.81** | 2.50 |
| | | 0.75 | 131 | 2.50 | **2.59** | 2.12 | 2.18 | **2.67** | 2.50 | 2.42 |
| | | 0.90 | 73 | **2.78** | **2.74** | 2.08 | **1.83** | **2.73** | 2.37 | 2.38 |
| | 2:1 | 0.25 | 212 | **3.20** | **2.91** | **2.60** | 1.81 | 2.36 | 2.27 | 2.45 |
| | | 0.40 | 254 | **2.73** | **2.65** | 2.23 | 2.07 | **2.52** | 2.47 | 2.47 |
| | | 0.60 | 240 | 2.31 | **2.69** | **1.92** | 2.30 | **2.74** | **2.62** | 2.41 |
| | | 0.75 | 180 | 2.21 | 2.50 | **1.74** | 2.18 | **2.78** | **2.61** | 2.50 |
| | | 0.90 | 88 | 2.02 | 2.41 | **1.21** | 2.27 | **2.92** | **2.83** | 2.41 |
| 0.05 | 1:2 | 0.95 | 289 | **3.41** | **2.81** | **2.82** | 1.76 | 2.17 | 2.17 | 2.43 |
| | 1:1 | 0.95 | 334 | **2.85** | **2.62** | 2.31 | **1.93** | 2.41 | 2.41 | 2.41 |
| | 2:1 | 0.95 | 414 | 2.14 | 2.49 | **1.56** | 2.14 | **2.90** | **2.55** | 2.40 |

Analysis of two Binomial Proportions in Non-inferiority Trials

Table 2: True Type 1 Errors for Various Tests for Studies with 90% Power at Nominal α=2.5%

| $\delta_0$ | $N_T:N_C$ | $P_C$ | $N_T$ | Wald | AC | HA | NCC | NC | ALS | ELS |
|---|---|---|---|---|---|---|---|---|---|---|
| 0.10 | 1:2 | 0.25 | 280 | 2.24 | 2.48 | **1.89** | 2.32 | **2.71** | **2.59** | 2.49 |
|  |  | 0.40 | 369 | 2.37 | 2.50 | 2.11 | 2.32 | **2.61** | **2.55** | 2.50 |
|  |  | 0.60 | 382 | 2.46 | 2.48 | 2.22 | 2.21 | 2.48 | 2.44 | 2.44 |
|  |  | 0.75 | 310 | **2.67** | **2.60** | 2.35 | 2.17 | 2.49 | 2.46 | 2.49 |
|  |  | 0.90 | 172 | **3.20** | **2.79** | **2.60** | 1.89 | 2.33 | 2.28 | 2.43 |
|  |  | 0.95 | 113 | **4.03** | **2.93** | **2.93** | 1.52 | 2.12 | 2.12 | 2.31 |
|  | 1:1 | 0.25 | 395 | **2.55** | **2.60** | 2.28 | 2.20 | **2.57** | 2.49 | 2.49 |
|  |  | 0.40 | 502 | 2.46 | **2.51** | 2.29 | 2.31 | **2.54** | 2.49 | 2.48 |
|  |  | 0.60 | 502 | **2.53** | **2.53** | 2.20 | 2.21 | **2.53** | **2.53** | **2.53** |
|  |  | 0.75 | 395 | **2.55** | **2.60** | 2.28 | 2.20 | **2.57** | 2.49 | 2.50 |
|  |  | 0.90 | 204 | **2.71** | **2.70** | 2.30 | 2.01 | **2.55** | 2.40 | 2.46 |
|  |  | 0.95 | 128 | **3.38** | **2.91** | 2.34 | 1.83 | 2.44 | 2.31 | 2.48 |
|  | 2:1 | 0.25 | 620 | **2.83** | **2.65** | 2.47 | 2.11 | 2.46 | 2.40 | 2.47 |
|  |  | 0.40 | 764 | **2.60** | **2.56** | 2.33 | 2.24 | **2.51** | 2.49 | 2.49 |
|  |  | 0.60 | 738 | 2.40 | **2.57** | 2.14 | 2.33 | **2.58** | **2.57** | **2.54** |
|  |  | 0.75 | 560 | 2.31 | 2.48 | 2.02 | 2.32 | **2.64** | **2.59** | 2.47 |
|  |  | 0.90 | 262 | 2.14 | **2.51** | **1.61** | 2.24 | **2.80** | **2.64** | 2.50 |
|  |  | 0.95 | 154 | 2.16 | **2.67** | **1.29** | 1.95 | **2.83** | **2.59** | 2.34 |
| 0.15 | 1:2 | 0.25 | 121 | 2.15 | **2.56** | **1.63** | 2.30 | **2.82** | **2.66** | 2.42 |
|  |  | 0.40 | 161 | 2.25 | **2.53** | **1.89** | 2.30 | **2.72** | **2.59** | 2.47 |
|  |  | 0.60 | 169 | 2.42 | 2.44 | 2.08 | 2.10 | **2.55** | 2.42 | 2.42 |
|  |  | 0.75 | 140 | **2.69** | **2.62** | 2.27 | 2.07 | **2.55** | 2.45 | 2.46 |
|  |  | 0.90 | 83 | **3.35** | **2.98** | **2.67** | 1.70 | 2.36 | 2.20 | 2.36 |
|  | 1:1 | 0.25 | 176 | **2.67** | **2.80** | 2.21 | 2.08 | **2.64** | 2.49 | 2.49 |
|  |  | 0.40 | 221 | 2.47 | **2.62** | 2.17 | 2.22 | **2.65** | **2.51** | 2.46 |
|  |  | 0.60 | 221 | 2.37 | 2.38 | 2.37 | 2.37 | 2.47 | 2.43 | 2.37 |
|  |  | 0.75 | 176 | 2.50 | **2.62** | 2.11 | 2.16 | **2.64** | **2.56** | 2.50 |
|  |  | 0.90 | 96 | **2.76** | **2.76** | 2.28 | 1.88 | **2.73** | 2.47 | 2.47 |
|  | 2:1 | 0.25 | 280 | **3.16** | **2.84** | **2.54** | 1.87 | 2.42 | 2.28 | 2.45 |
|  |  | 0.40 | 338 | **2.67** | **2.63** | 2.25 | 2.14 | **2.54** | 2.49 | 2.49 |
|  |  | 0.60 | 322 | 2.31 | **2.63** | **1.95** | 2.26 | **2.63** | **2.63** | **2.58** |
|  |  | 0.75 | 242 | 2.22 | 2.50 | **1.81** | 2.26 | **2.76** | **2.62** | 2.50 |
|  |  | 0.90 | 120 | **1.98** | 2.40 | **1.30** | 2.34 | **2.94** | **2.84** | 2.40 |
| 0.05 | 1:2 | 0.95 | 374 | **3.30** | **2.77** | **2.77** | 1.82 | 2.24 | 2.23 | 2.47 |
| 0.05 | 1:1 | 0.95 | 440 | **2.82** | **2.65** | 2.35 | 2.02 | 2.48 | 2.37 | 2.48 |
| 0.05 | 2:1 | 0.95 | 560 | 2.12 | 2.43 | **1.62** | 2.18 | **2.78** | **2.62** | 2.44 |

Analysis of two Binomial Proportions in Non-inferiority TrialsTable 3: Summary of the Type I Errors for the Various Scenarios Presented in Tables 1 and 2

| Type I Errors | Wald | AC | HA | NCC | NC | ALS | ELS |
|---|---|---|---|---|---|---|---|
| % Of type I errors >2.5% | 49% | 74% | 14% | 0 | 69% | 42% | 8% |
| Mean Distance from 2.5% | 0.324 | 0.145 | 0.410 | 0.401 | 0.161 | 0.120 | 0.046 |
| Range | 2.05 | 0.7 | 1.84 | 0.96 | 0.92 | 0.79 | 0.27 |
| Min | 1.98 | 2.38 | 1.21 | 1.41 | 2.05 | 2.05 | 2.31 |
| Max | 4.03 | 3.08 | 3.05 | 2.37 | 2.97 | 2.84 | 2.58 |
| Mean of those ≤2.5% | 2.30 | 2.46 | 2.06 | 2.10 | 2.37 | 2.38 | 2.45 |
| Mean of those >2.5% | 2.95 | 2.68 | 2.71 | -- | 2.67 | 2.62 | 2.53 |

Mean distance from 2.5% is defined as the average absolute value of the difference of type I errors from the 2.5% nominal value across the 72 scenarios displayed in Tables 1 and 2.

Analysis of two Binomial Proportions in Non-inferiority Trials

Table 4. p-values and 95% CIs for testing non-inferiority hypotheses for 3 selected examples

| Data: | | Example 1 | | Example 2 | | Example 3 | |
|---|---|---|---|---|---|---|---|
| Test: | $x_T / N_T$ (%) | 264/328 (80.5%) | | 285/326 (87.4%) | | 411/435 (94.5%) | |
| Control: | $x_C / N_C$ (%) | 268/317 (84.5%) | | 99/108 (91.7%) | | 426/441 (96.6%) | |
| Null Margin: | $\delta_0$ | 10% | | 10% | | 5% | |
| Method | | p-value | (95% CI) | p-value | (95% CI) | p-value | (95% CI) |
| Wald | | -- | **(-9.91, 1.80)** | -- | (-10.58, 2.09) | -- | **(-4.85, 0.62)** |
| AC | | -- | **(-9.88, 1.84)** | -- | (-10.19, 2.76) | -- | **(-4.89, 0.68)** |
| HA | | -- | (-10.07, 1.96) | -- | (-11.06, 2.58) | -- | (-4.97, 0.73) |
| NC | | -- | **(-9.90, 1.83)** | -- | **(-9.85, 3.21)** | -- | (-5.00, 0.66) |
| NCC | | -- | (-10.11, 2.06) | -- | (-10.20, 3.78) | -- | (-5.16, 0.83) |
| ALS | | 0.0238 | (-9.94, 1.83) | **0.0246** | **(-9.98, 3.16)** | 0.0260 | (-5.03, 0.64) |
| ELS | | 0.0239 | (-9.94, 1.84) | 0.0281 | (-10.14, 2.91) | 0.0246 | (-4.99, 0.66) |
| Real computer time to obtain the CI & p-value for the ELS method | | 12 seconds | | 5 seconds | | 23 seconds | |

p-values are for non-inferiority comparison of test to control groups and only generated for the test-based confidence interval methods; CI: confidence interval;
**Bolded** data indicate methods with type I errors likely being higher than the nominal value of 2.5% based on the true proportions being close to the considered observed proportions.